\documentclass[a4paper,10pt]{llncs}
\usepackage[utf8]{inputenc}
\usepackage[english]{babel} 
\usepackage{epsfig}
\usepackage{amsmath}
\usepackage{amssymb}

\author{Toni Heidenreich}
\institute{King's College London}
\title{The formal-logical characterisation of lies, deception, and associated notions}

\begin{document}

\maketitle

\abstract{Defining various dishonest notions in a formal way is a key step to enable intelligent agents to act in untrustworthy environments. This review evaluates the literature for this topic by looking at formal definitions based on modal logic as well as other formal approaches. Criteria from philosophical groundwork is used to assess the definitions for correctness and completeness. The key contribution of this review is to show that only a few definitions fully comply with this gold standard and to point out the missing steps towards a successful application of these definitions in an actual agent environment.}
\setlength{\parindent}{1.5em}

\section{Introduction}

A formal-logical characterisation of deception and associated notions can be approached with different methods. This review considers two principal methods: formal definitions based on pure modal logic as well as definitions based on agent architecture and communication. We will evaluate how the various formal definitions comply with the philosophical foundations of this topic. Moreover, the respective advantages and disadvantages of these approaches are extracted and used to determine open issues for further research. Putting it most simply, two questions will be answered: Are the existing formal-logical definitions correct given the philosophical background? And what are the problems needing further attention?

Deception has been part of Computer Science since its very beginning. The famous "Imitation Game" proposed by Alan Turing to test whether machines can think \cite{Turing1950} explicitly asked for a machine being able to deceive the interrogator.  
In more practical terms the topic has not been studied until recently when computers became more connected and the concept of agents and multi-agent systems emerged. In 2003, the \textit{AgentLink} community identified trust in agent systems as one of the key challenges for further research \cite{Luck2003}. This included user confidence in trustworthiness of machines as well as trust in norms and social rules among agents. As Jones and Firozabadi pointed out \cite{Jones2001}, trust in this context means trust in the reliability to tell the truth as opposed to other notions like trust in the abilities of others. Dishonesties in this domain have been cause for concern, especially in such application areas as automatic trading agents (e.g. \cite{Solomon2013}), automatic negotiations (e.g. \cite{Rosenschein1994}) or trust in the security of other computer systems (e.g. \cite{Stech2011,Barber2003}). In particular, in open multi-agent systems where agents are free to leave or enter the system, trust cannot be taken for granted as, e.g., argued in \cite{Stranders2006}.

Since it seems to be established that trust is an important topic in computer science and multi-agent systems in particular, a lot of authors defined trust in a logical-formal way \cite{Demolombe2004,Grandison2000,Huang2010}. This review, however, will focus on the opposite: definitions of not telling the truth, of lying, being deceived or related notions of dishonesty. Although being closely connected, remarkably little work has been done to define these concepts formally, as indicated by many of the authors in the field \cite{ONeill2003,Tzouvaras1998,Sakama2010a,Pan2007,Rahwan2008}. Summing up, it can be shown that trust is an important topic which has been formally defined by many authors whereas the associated negative notions evolving around lies and deception received comparatively little attention.

In the following chapter 2, we will look at the philosophical work to identify the necessary criteria that definitions of lying and deception need to satisfy in order to be accepted. Chapter 3 will then review some of the existing definitions based on modal logic. Chapter 4 reviews some formal definitions based on other approaches. Finally, chapter 5 will summarise and compare in order to determine open research topics of the area. 

\section{Philosophical foundation}

Many of the papers introduced later base their work on philosophical definitions of lying and deception. Even more important, philosophy has discussed dishonesties for far longer than computer scientists and logicians. Thereby it enables us to identify conditions and criteria that need to be fulfilled by a definition to capture the underlying concepts correctly. 

Unfortunately, no definition is universally accepted as different authors indicated \cite{Mahon2008b,Sakama2011}. Almost every definition can be disputed to encompass either not all or too many cases. Moreover, the boundary between lies and deception is not unambiguous \cite{Adler1997}. As recent as 2010, it is claimed that philosophers argue about the right way to define these notions \cite{Fallis2010}. Despite the ambiguity, we will try to find the conditions supported by the majority of philosophers. In the first section this will be done for the notion of lying, in the second section for deception and the third section will cover other similar notions defined by philosophers. To illustrate the various concepts, the running example of an estate agent guiding a customer through a flat will be used. In this scenario, lying and deceiving may occur very naturally making the example easier to understand.

\subsubsection{Lying}

We will identify five conditions necessary to describe what constitutes a lie. These conditions are based on those expatiated in the Stanford Encyclopedia of Philosophy \cite{Mahon2008b}. 

Starting off with a dictionary definition, the Cambridge Dictionary says lying is "to say or write something which is not true in order to deceive someone" \cite{Cambridge2008}. This definition includes the first condition, the \textit{statement condition}, meaning that a lie is only a lie if a spoken statement or utterance occurred (defined e.g. by \cite{Chisholm1977}). In the example, the estate agent is therefore lying when he says "I've been in the business for more than 10 years", although he knows he just started. He is, however, not lying when he behaves and dresses himself as if he is an expert. Some philosophers argue against this condition \cite{Vrij2000} maintaining a broader view that even withholding helpful information should be considered a lie.

What the dictionary definition is probably doing wrong, is to assert that the utterance must be false. This is best illustrated with the case where the estate agent honestly believes that the flat has been renovated recently although it has not. According to the definition, he would be lying by telling the customer that the flat has been renovated recently, although it is his honest believe. To exclude this dilemma, most philosophers introduce the \textit{believe-false condition}, where the liar must believe the proposition he is saying to be false, independent of the actual truth-value. This is reflected, for example, in the definition of J.\,Kupfer that "a person lies when he asserts something to another which he believes to be false with the intention of getting the other to believe it to be true" \cite{Kupfer1982}. Similar definitions have been given by I.\,Primoratz \cite{Primoratz1984} or B.\,Williams \cite{Williams2004}. Some like Chisholm and Feehan \cite{Chisholm1977} argue that a \textit{not-believe-true condition} is also enough which includes the case where the liar has no believe at all about what he is saying.

Kupfer's definition is also clarifying the role of the listener. First of all there must be a listener, usually called the \textit{addressee condition} \cite{Mahon2008b}. Thereby cases of telling some wrong believe to an empty room or being eavesdropped telling a wrong belief are not lies per se. Furthermore, the definition highlights the \textit{intent-to-deceive condition} which has also been described by J.\,Mahon \cite{Mahon2008a}. Our estate agent joking about the value of the property by telling an exorbitant price is not lying, because he has no intent to deceive the customer with this statement. Even if some like R.\,Sorenson \cite{Sorensen2007} or T.\,Carson \cite{Carson2010} argue against this condition in special cases, general adaptations should include this condition to rule out fakes, jokes or play-acts \cite{Fallis2010}.

The fifth condition is the demand that a definition should in no case include the \textit{success condition}, meaning that the intended deception was successful \cite{Mahon2008b}. Imagine the case when the estate agent lies about the quality of the parquet and the customer happens to be a skilled carpenter being able to judge the quality correctly. In this occasion he would still think that the estate agent lied, so the success (or failure) of the attempted deception makes no difference to the fact that a lie occurred.

It is interesting to note that none of the mentioned conditions state that a lie is morally wrong \cite{Mahon2008b}. As reasoned by Kemp and Sullivan this is because morality is "a synthetic judgement and not an analytic one" \cite{Kemp1993}. There are also other possible conditions for lying which are, for example, defined by C.\,Sakama in \cite{Sakama2011}. Since he also gives formal definitions, we won't use his conditions to ensure that the standard is not biased towards his definition.

This leaves us with the five conditions from the Stanford Encyclopedia of Philosophy that a lie is only correctly defined if it contains the statement condition, the believe-false-condition, the addressee condition, the intent-to-deceive condition and not the success-condition or any other restrictions.

\subsubsection{Deceiving}

The Cambridge Dictionary defines deceiving as "persuade someone that something false is the truth" \cite{Cambridge2008}. This definition shows us that deception is concerned with the effect on the listener or receiver of the message as opposed to lying which is focused on the dishonest behaviour of the speaker. For this reason, three changes need to be made to the list of necessary conditions.

As already indicated, the first change is to include the \textit{success condition}. When the estate agent lied about the parquet, he did not manage to deceive the carpenter. Deception would have only occurred if the customer had believed the lie. As argued by Chisholm and Feehan in \cite{Chisholm1977}, the proposition the listener believes after the deception does not necessarily need to be a new belief. He could also be deceived by maintaining a belief or even by being prevented from acquiring some belief (in this case the success is that he continues to believe that the proposition is not true).

The second change is to exclude the \textit{statement condition} as reasoned by J.\,Mahon \cite{Mahon2008b} or L.\,Linsky \cite{Linsky1963}. Coming back to a previous example, a knowledgeable appearance of the estate agent does not constitute a lie. But if this is intentional, the customer could be deceived about his expertise without any statement being made. Instead, a new criterion called \textit{evidence condition} is added. It says that the deceiving person must provide some form of evidence which is the reason for the listener to conclude the wrong proposition \cite{Fuller1976,Barnes2007,Mahon2007}. This ensures the agency of the deceiving person. For example, if a friend of the customer told him that the estate agent has a high level of expertise and the agent still wears the potentially misleading outfit and the belief of the customer is solely based on the friend, then the estate agent has definitely not deceived the customer. In another example, the estate agent might even tell the truth by saying that the parquet is shining brightly with the intention that the customer concludes a high quality. Still, this is a form of deception as evidence intentionally makes the customer believe something that the estate agent does not believe.

All other conditions still hold for deception: the \textit{believe-false} or its weaker form   \textit{not-believe-true condition}, the \textit{addressee condition} and the \textit{intent-to-deceive condition} (e.g. \cite{Mahon2007}). Some like J.\,Adler argue that deception does not need to be intentionally \cite{Adler1997}. But as most of the other authors disagree with this position we will still use it as a necessary condition.

As a result, we know that a definition of deception needs to fulfil the believe-false condition, the addressee condition, the intent-to-deceive condition, the evidence condition and the success condition. It must not include the statement condition or other restricting notions.

\subsubsection{Other notions}

The definitions of dishonest behaviours given by philosophers are not limited to lies and deception, although being the main focus. Other notions which have been considered are fraud, bullshit, withholding information or half-truths. They became necessary as the available definitions of lying and deceiving did not include all kinds of possible dishonesties.

M.\,Simmons defined in \cite{Simmons1995} the notion of \textit{fraud}. His definition essentially includes a lie which is believed by the listener and therefore also becomes a successful deception, combined with requirement that the victim uses the acquired information and makes loss of money or property as a result. This definition shall not be used as a guideline in this review but is given here because one of the formal definitions refers to it.

A more relevant notion is \textit{bullshit} as it was denoted by H.\,Frankfurt in \cite{Frankfurt2005}. Bullshit covers similar cases to lying with the difference that the speaker does not follow the believe-false condition and instead neither believes the statement to be true nor does he believe it to be false. The estate agent bullshits, for instance, when he highlights the satisfaction of the previous tenants even though he did not know them.

Another important notion is \textit{withholding information} as e.g. mentioned by T.\,Carson in \cite{Carson2010}. It is the failure to offer information that would help to acquire true beliefs or correct false beliefs as long as this result is intentional. For example, not telling that the previous tenant moved out because of noisy neighbours is an obvious example of withholding information. In this case, the conditions for lies can be adopted as well, with the difference that the statement condition is replaced by the explicit non-statement that would have helped to change the belief of the listener. Carson also mentions \textit{half-truths} in the same paper as a special case of deception including a true statement as evidence.

Although a number of other notions have been defined, mostly bullshit and withholding information will be relevant for the formal definitions.

Altogether, this chapter provided a basic overview of how philosophers define lies, deception and similar concepts. For lies and deception we were able to extract a concrete list of conditions allowing us to check whether formal definitions comply with this standard.

\section{Modal logic definitions}

This chapter will look at definitions of lies and deception in modal logic. Using logic to describe these concepts is an obvious decision as it provides a very general expressiveness which is well understood and not restrained to an application area. However, as standard propositional logic is not enough to capture all the subtle notions, almost all definitions make use of some kind of modal logic. By introducing additional modal operators, this kind of logic allows for quantifications not possible with simple true/false values (cf. Stanford Encyclopedia of Philosophy \cite{Garson2013}). Different authors use different sets of operators; for simplicity we will introduce the five most common and explain less common operators later when they are applied:

\begin{itemize}
\item $\mathcal{B}_ip$ denotes that agent $i$ \textit{believes} the fact $p$. This does not say anything about the underlying truth of $p$ and just indicates the beliefs of agent $i$.  This operator and the following two often occur together in a system called $\mathcal{BIC}$ which was e.g. specified by M.\,Colombetti \cite{Colombetti1999}.
\item $\mathcal{I}_ip$ denotes that agent $i$ has the \textit{intention} to make $p$ true.  For example, by writing $\mathcal{I}_i\mathcal{B}_jp$ we can specify that the estate agent $i$ has the intention that the customer $j$ believes that the flat is renovated (=$p$). 
\item $\mathcal{C}_{ij}p$ denotes that agent $i$ \textit{communicates} the fact $p$ to agent $j$. This includes any kind of spoken communication or utterance.
\item $\mathcal{O}p$ denotes that it \textit{ought to be} $p$, where $p$ is any proposition. The operator is e.g. described by B.\,Chellas \cite{Chellas1980} or A.\,Jones \cite{Jones1985} and means that in known environments it is ideally the case that $p$. For example in the flat sale situation, it ought to be the case that the estate agent is authorised to sell the flat.
\item $\mathcal{E}_ip$ denotes that an agent $i$ \textit{brings about} that $p$ and was introduced by I.\,Pörn \cite{Porn1977}. It assigns the agency of $i$ to the fact $p$, in the sense that agent $i$ is the decisive factor that $p$ occurred or became true. Then a friend of the customer told him that the estate agent is an expert in his field, this friend brought about that the customer believes in the expertise.
\end{itemize} 
Authors of papers using these operators usually provide proofs or refer to others who proofed that the logical system is coherent and fulfils a number of desired properties. We won't go into details at this point as interested readers may refer to the original papers.

In the following, we will start to look at formal definitions of lying before continuing with bullshit, deception and other notions. Each considered paper is evaluated using the conditions selected in the last chapter. Within each group a chronological order will be maintained if reasonable. Thus, developments and improvements over time can be seen.

\subsubsection{Lying}

\paragraph{B.\,O'Neill (2003) \cite{ONeill2003}.} 

The first definition we will examine is one given by B.\,O'Neill. In his paper, he defines and derives several properties of the modal operator $\mathcal{C}$ for communication and subsequently defines lying with the formula in equation \ref{on2003_1}. Starting from his previously derived properties of communication he then shows that  lying is a subset of all situations which satisfy equation \ref{on2003_2}.
\begin{gather}
\label{on2003_1}
\mathcal{C}_{ij}p\wedge\mathcal{B}_i\neg p \\
\label{on2003_2}
\mathcal{I}_i\mathcal{B}_j\mathcal{B}_ip\wedge\mathcal{B}_i\neg\mathcal{B}_ip
\end{gather}
In an example, this might mean that the estate agent $i$ is lying when he tells customer $j$ that the flat has been renovated ($\mathcal{C}_{ij}p$) while believing this is not true ($\mathcal{B}_i\neg p$). As a matter of some rules governing communication it implies that he intends the customer to believe that he believes the flat has been renovated ($\mathcal{I}_i\mathcal{B}_j\mathcal{B}_ip$) and that he does not believe that the renovation is part of his belief ($\mathcal{B}_i\neg\mathcal{B}_ip$). 

It can be seen very easily that the statement condition ($\mathcal{C}_{ij}p$), the addressee condition ($j$) and the believe-false condition ($\mathcal{B}_i\neg p$) are fulfilled. The definition also contains no notion of success and some intention. The problem is, however, that this intention is not exactly the desired intent-to-deceive condition as agent $i$ is not intending that $j$ actually believes $p$, but rather that $j$ accepts that $i$ is telling his true belief ($\mathcal{I}_i\mathcal{B}_j\mathcal{B}_ip$). By adding this extra level of abstraction, O'Neill fails to meet this condition.

On the other hand he shows very well that besides the believe-false condition, the weaker not-believe true condition can constitute some form of dishonesty which he calls 'talking though one's hat'. The according formal definition just replaces $\mathcal{B}_i\neg p$ with $\neg\mathcal{B}_ip$ in equation \ref{on2003_1}.

As a result, we can say that O'Neill gives a well-thought-of definition with a small inconsistency and manages to relate different levels of lying in the formal definition.

\paragraph{M.\,Caminada (2009) \cite{Caminada2009}.}

This paper of Caminada mainly focuses on the difference between lying and bullshit. In this context he defines lying as given in equation \ref{ca2009}. In parts, this definition is very similar to one given by A.\,Tzouvaras 11 years earlier \cite{Tzouvaras1998}, but is more clarified by leaving out unnecessary parts.
\begin{equation} \label{ca2009}
\mathcal{C}_ip\wedge\mathcal{B}_i\neg p
\end{equation}
His definition complies with the statement condition ($\mathcal{C}_ip$), the believe-false condition ($\mathcal{B}_i\neg p$) and it contains no notion of success. However, it suffers from the problem of not including anything concerned with the listener, so that neither the addressee condition nor the intent-to-deceive conditions are satisfied.

Despite the lacking expressiveness of this definition he claims that the definition of lying is settled and well-defined. He mentions, however, that the intent-to-deceive condition could be added but argues that this easier approach is sufficient. Nevertheless, the formal definition as it was given should be rejected for the named reasons.

\paragraph{C.\,Sakama et al. (2010) \cite{Sakama2010a}.}
In this paper the authors try to formally define lying, as well as bullshit and deception. The other   definitions besides lying will be given later for clarity reasons. 

Using a similar logical framework as all aforementioned authors, their definition of lying in equation \ref{sa2010} is the first satisfying all the conditions.
\begin{equation} \label{sa2010}
\mathcal{C}_{ij}p\wedge\mathcal{B}_i\neg p\wedge\mathcal{I}_i\mathcal{B}_jp
\end{equation}
It contains a statement ($\mathcal{C}_{ij}p$) to an addressee, the speaker fulfils the believe-false condition ($\mathcal{B}_i\neg p$), the intent-to-deceive condition ($\mathcal{I}_i\mathcal{B}_jp$) and no notion of success is included. The definition is therefore fully compatible with the philosophical criteria.

In addition, the authors give more specialised versions of this definition which include the objective of the liar. They also conclude that lies have to be as weak as possible to deceive the listener as they always introduce some deviation from the truth (or from what the liar believes to be the truth). That later binds him to his lie and makes him less free in what he can say without contradicting himself. This observation is quiet application-oriented and shows more insight than other papers. All in all, this definition shows to be the most comprehensive definition among those which have been reviewed, both in accuracy and profundity.
\newpage
\subsubsection{Bullshit}

\paragraph{M.\,Caminada (2009) \cite{Caminada2009}.} The first and oldest formal definition of bullshit is the one given in Caminada's paper which already included a definition of lying. Since bullshit as a philosophical concept was only defined in 2005 by H. Frankfurt \cite{Frankfurt2005} this is just one of two available formal definitions. Bullshit in Caminada's definition in equation \ref{ca2009_bs} highlights the main difference to lying: that the speaker has neither a belief that his statement is true ($\neg\mathcal{B}_ip$) nor that it is false ($\neg\mathcal{B}_i\neg p$).
\begin{equation} \label{ca2009_bs}
\mathcal{C}_ip\wedge\neg\mathcal{B}_ip\wedge\neg\mathcal{B}_i\neg p
\end{equation}
Similar to his definition of lying, it also contains a statement ($\mathcal{C}_ip$), but no addressee and no intention to deceive. In contrast to lying, this is not necessarily a problem as Frankfurt's informal definition does not specifically contain these parts either. Lending the example given by C.\,Sakama in \cite{Sakama2010a}, we can imagine that the estate agent is providing a consulting service and is paid per hour or report length. This might cause him to produce some bullshit according to equation \ref{ca2009_bs} just to earn more money but without any intent that the reader is actually believing what he has written, since it makes no difference as long as he appears knowledgeable.

\paragraph{C.\,Sakama et al. (2010) \cite{Sakama2010a}.}

Sakama et al. present bullshit as a weaker form of lying in the already mentioned paper. They actually give exactly the same definition as Caminada (Eq. \ref{ca2009_bs}). Furthermore, they produce a definition for 'intentional bullshit' (Eq. \ref{sa2010_bs}) including the missing intent-to-deceive condition ($\mathcal{I}_i\mathcal{B}_jp$) and the addressee condition.
\begin{equation} \label{sa2010_bs}
\mathcal{C}_{ij}p\wedge\neg\mathcal{B}_ip\wedge\neg\mathcal{B}_i\neg p\wedge\mathcal{I}_i\mathcal{B}_jp
\end{equation}
Relating to their consideration about the impact lying may have on future communication, they conclude that intentional bullshit should be preferred over lying, if possible, as it does not deviate from the true belief as much as a lie. This conclusion seems quite natural as people are more likely to tolerate bullshit than lies as H.\,Frankfurt pointed out in his work \cite{Frankfurt2005}.

\subsubsection{Deception}

\paragraph{B.\,Firozabadi et al. (1998) \cite{Firozabadi1998}.}
This paper by Firozabadi et al. is focused on verifying trade procedures by excluding fraud. Based on the definition of fraud given by M.\,Simmons \cite{Simmons1995} they produce four different possible formal definitions of deception (which is a constituting part of fraud). In contrast to all other authors we looked at so far, they use the modal operators $\mathcal{B}$ for belief, $\mathcal{E}$ for 'brings about' and a derived operator $\mathcal{H}$ for 'attempts to bring about'. The latter was first introduced by F.\,Santos et al. \cite{Santos1997} and has the same meaning as 'bringing about' something but without the inherent success of the action.
\begin{gather}
\label{fi1998_1}
\neg\mathcal{B}_ip\wedge\mathcal{E}_i\mathcal{B}_jp \\
\label{fi1998_2}
\neg\mathcal{B}_ip\wedge\mathcal{H}_i\mathcal{B}_jp \\
\label{fi1998_3}
\mathcal{B}_i\neg p\wedge\mathcal{E}_i\mathcal{B}_jp \\
\label{fi1998_4}
\mathcal{B}_i\neg p\wedge\mathcal{H}_i\mathcal{B}_jp
\end{gather}
In the deception definitions (Eq. \ref{fi1998_1} to \ref{fi1998_4}) they include the believe-false ($\mathcal{B}_i\neg p$) or the weaker not-believe-true condition ($\neg\mathcal{B}_ip$) and the deception is directed at an addressee ($j$) without the notion of communication. Both operators, $\mathcal{E}$ and $\mathcal{H}$, denote the agency of $i$ and do not include any intention. Therefore, the intent-to-deceive condition is not included. Moreover, the  $\mathcal{H}$ operator does not even include the success of the deception, that is why equations \ref{fi1998_2} and \ref{fi1998_4} need to be rejected. The last criterion asking for evidence is not included as well. In summary, these definitions lack several of the necessary conditions and are not up to the standard given by the philosophical literature.

\paragraph{A.\,Jones et al. (2001) \cite{Jones2001}.}  

In this paper A.\,Jones and B.\,Firozabadi (the author of the previous paper) improve on the definition of deception. Again, they use the belief operator $\mathcal{B}$ and the operator $\mathcal{E}$ for 'bringing about' something. Additionally, they use the already introduced operator $\mathcal{O}$ for 'it ought to be that' and another operator $a\Rightarrow_sb$ denoting that $a$ 'counts as' $b$ given the context or institutionalised power of $s$. Simplified, the operator which was introduced by A.\,Jones in \cite{Jones1996} denotes a consequence which is true under given circumstances.  For example, getting a plastic card with one's name on it ($a$) 'counts as' being a student ($b$) as long as this is done by an university ($s$).

Their definition of deception in equation \ref{jo2001} fulfils all conditions except the intent-to-deceive condition.
\begin{equation} \label{jo2001}
\neg\mathcal{B}_ip\wedge\mathcal{E}_i\mathcal{B}_j\mathcal{E}_im\wedge(((\mathcal{E}_im \Rightarrow_s\mathcal{O}p)\wedge\mathcal{B}_j\mathcal{E}_im)\rightarrow\mathcal{B}_jp)
\end{equation}
It doesn't contain a statement, but an addressee ($j$) and it complies with the not-believe-true condition ($\neg\mathcal{B}_ip$). The evidence condition is included ($\mathcal{E}_i\mathcal{B}_j\mathcal{E}_im$ and $\mathcal{E}_im \Rightarrow_s\mathcal{O}p$) in a way that $i$ brings about that $j$ believes he brought about the evidence $m$, while under the current circumstances $s$ this bringing about of evidence $m$ ought to mean that $p$ is true. The success of the deception is included as well, as this allows $j$ to reason that $p$ is true ($\rightarrow\mathcal{B}_jp$). The missing intention to deceive is, however, only a minor problem as this was one of the disputed conditions for deception anyway (as mentioned in chapter 2).

\paragraph{G.\,Meggle (2000) \cite{Meggle2000}.}
Unlike the previous authors, Meggle uses a variant of the $\mathcal{BIC}$ logic to describe deception and associated notions. 

\begin{equation} \label{me2000}
\mathcal{I}_i\mathcal{B}_jp\wedge\mathcal{B}_i\neg p\wedge\mathcal{C}_{ij}m\wedge\mathcal{B}_i(\mathcal{C}_{ij}m\rightarrow\mathcal{B}_jp)
\wedge(\mathcal{C}_{ij}m\rightarrow\mathcal{B}_jp)
\end{equation}
The definition he gives for successful deception in equation \ref{me2000} correctly contains the intent-to-deceive ($\mathcal{I}_i\mathcal{B}_jp$) and the believe-false condition ($\mathcal{B}_i\neg p$). It provides evidence ($\mathcal{C}_{ij}m$) which $i$ believes to cause $j$ to acquire the new belief ($\mathcal{B}_i(\mathcal{C}_{ij}m\rightarrow\mathcal{B}_jp)$) and it also covers the success in the way that this evidence indeed causes $j$ to acquire the new belief ($\mathcal{C}_{ij}m\rightarrow\mathcal{B}_jp$). The problem of this definition is, however, that it explicitly contains a statement condition which should be avoided to account for deception by other means.

Furthermore, Meggle uses his definition to nest multiple levels of deception (being deceived about being deceived and so on) and thinks about implications for actual implementation. These considerations and the good, albeit not perfect, definition contribute to make this a valuable paper.

\paragraph{B.\,O'Neill (2003) \cite{ONeill2003}.} 
In the already mentioned paper of B.\,O'Neill, he also gives a definition of deception as in equation \ref{on2003_de}. In his paper he actually starts by defining deception, only to show later that his definition of lying is a subset of deception.
\begin{equation} \label{on2003_de}
\mathcal{I}_i\mathcal{B}_jp\wedge\mathcal{B}_i\neg p\wedge\mathcal{B}_jp
\end{equation}
He correctly leaves out a statement, but has an addressee ($j$), fulfils the believe-false condition ($\mathcal{B}_i\neg p$), the intent-to-deceive condition ($\mathcal{I}_i\mathcal{B}_jp$) and the success condition ($\mathcal{B}_jp$). Again, one condition is not fulfilled, as no trace of an evidence is included in his definition.

\paragraph{C.\,Sakama et al. (2010) \cite{Sakama2010a}} In the same paper where they already defined lies and bullshit, Sakama et al. also define deception with equation \ref{sa2010_de}. They try to focus on the speaker's point of view, thereby neglecting some of the necessary conditions.

\begin{equation} \label{sa2010_de}
\mathcal{C}_{ij}m\wedge\mathcal{B}_im\wedge\mathcal{I}_i\mathcal{B}_jm\wedge\mathcal{B}_i \mathcal{B}_j(m\wedge\neg \mathcal{B}_j\neg p\rightarrow p)\wedge\mathcal{B}_i\neg\mathcal{B}_j\neg p\wedge\mathcal{B}_i\neg p\wedge\mathcal{I}_i\mathcal{B}_jp
\end{equation}
The way to interpret this rather long statement is that $i$ communicates some evidence $m$ which he believes himself and intents $j$ to believe it as well. He furthermore thinks that $j$ makes the default conclusion that $p$ holds as well (as long as he does not believe that $\neg p$). Agent $i$ also expects that $j$ does not believe $\neg p$ while he himself does not believe in $p$ with the overall intention the $j$ comes to believe $p$.

It contains an addressee, the believe-false condition, the intent-to-deceive and some evidence. The problems are that it does not state at any point that the deception was successful and that it introduces unnecessary restrictions by using the statement condition and restricting the evidence to propositions he believes himself ($\mathcal{B}_im$).

Even though the definition lacks several of our criteria, the authors do well in concluding that deception is even better than bullshit or lying as it does not even require to deviate from the truth at all.

\paragraph{C.\,Sakama et al. (2010) \cite{Sakama2010b}.} In this paper of C.\,Sakama and M.\,Caminada, the authors try a different approach to define deception by enumerating all possible situations which might constitute deception. They base these definitions on the work of Chisholm and Feehan \cite{Chisholm1977} who differentiated deception by aim, effect and knowledge. Besides the familiar operators of belief ($\mathcal{B}$), intention ($\mathcal{I}$), communication ($\mathcal{C}$) and bringing about something ($\mathcal{E}$), they additionally use 'let it be the case that' denoted by $\mathcal{F}$. This operator has a similar meaning to $\mathcal{E}$, but with the difference that $\mathcal{E}_ip$ denotes that the agency of $i$ changes $p$, while $\mathcal{F}_ip$ means that the agency of $i$ allows $p$ to continue to be what it was before.

The definitions in equations \ref{sa2010_81} to \ref{sa2010_88} can be summarised as telling or not telling something which causes the listener to start believing, continue believing, cease not believing or being prevented from not believing some proposition.
\begin{align}
\label{sa2010_81}
\mathcal{B}_i\neg p\wedge\mathcal{C}_{ij}p&\rightarrow\mathcal{E}_i\mathcal{B}_jp \\
\mathcal{B}_i\neg p\wedge\mathcal{C}_{ij}p&\rightarrow\mathcal{F}_i\mathcal{B}_jp \\
\mathcal{B}_i\neg p\wedge\mathcal{C}_{ij}p&\rightarrow\mathcal{E}_i\neg\mathcal{B}_j\neg p \\
\mathcal{B}_i\neg p\wedge\mathcal{C}_{ij}p&\rightarrow\mathcal{F}_i\neg\mathcal{B}_j\neg p \\
\mathcal{B}_i\neg p\wedge\neg\mathcal{C}_{ij}\neg p&\rightarrow\mathcal{E}_i\mathcal{B}_jp \\
\mathcal{B}_i\neg p\wedge\neg\mathcal{C}_{ij}\neg p&\rightarrow\mathcal{F}_i\mathcal{B}_jp \\
\mathcal{B}_i\neg p\wedge\neg\mathcal{C}_{ij}\neg p&\rightarrow\mathcal{E}_i\neg\mathcal{B}_j\neg p \\
\label{sa2010_88}
\mathcal{B}_i\neg p\wedge\neg\mathcal{C}_{ij}\neg p&\rightarrow\mathcal{F}_i\neg\mathcal{B}_j\neg p
\end{align}
The eight definitions exist without or with the intentional part $\mathcal{I}_i\mathcal{B}_jp$. By this they include the intent-to-deceive condition while not ruling out the possibility of unintentional deception. They clearly also have an addressee and a believe-false condition ($\mathcal{B}_i\neg p$). Furthermore all possible forms of success are included. At first it might seem problematic to include the statement condition ($\mathcal{C}_{ij}p$), but by enumerating the same equations with the explicit non-statement ($\neg\mathcal{C}_{ij}\neg p$) they actually show indifference to the statement as required by the philosophical criteria. The only thing that is definitely missing is the evidence condition.

All together, this definition stands out in its achievement to include all the different notions which might be included in deception with the downside that the evidence condition is missing completely.

\subsubsection{Other notions}

\paragraph{C.\,Sakama et al. (2010) \cite{Sakama2010b}.} 
Other notions of dishonesty defined in a formal way were only embedded in papers of C.\,Sakama. One of them is 'withholding information' which is defined in the same way as lying but with the non-statement instead of the statement condition (Eq. \ref{sa2010_wi}).
\begin{equation} \label{sa2010_wi}
\neg\mathcal{C}_{ij}\neg p\wedge\mathcal{B}_i\neg p\wedge\mathcal{I}_i\mathcal{B}_jp
\end{equation}
This definition can be considered as complete, since the non-statement condition, the addressee condition, the believe-false condition and the intent-to-deceive condition are included without using any success condition.

\paragraph{C.\,Sakama (2012) \cite{Sakama2012a}.} This presentation of Sakama includes a number of his previously mentioned definitions and additionally a definition of half-truths. Interestingly, half-truths have the same definition as his deception definition in \cite{Sakama2010a}, given in equation \ref{sa2010_de}. Presumably, he noticed the shortcomings of this definition as deception and relabelled it as half-truth which is indeed more appropriate given, e.g., the dictionary definition that a half-truth is "a statement that is intended to deceive by being only partly true" \cite{Cambridge2008}.

\subsubsection{Summary}

\paragraph{}After looking at a number of definitions in modal logic, we can conclude that fulfilling the various philosophical criteria is not at all obvious. For lying, only Sakama's definition in \cite{Sakama2010a} is fully compatible with the criteria. For deception, none of the reviewed papers included a definition which is totally correct. The definitions of Jones \cite{Jones2001}, Meggle \cite{Meggle2000} and Sakama \cite{Sakama2010b}, however, are the most suitable with only one condition missing in each definition. The latter manages to provide very subtle differences by using multiple statements with the downside of introducing possibly unnecessary complications. On the other hand, the definitions of the minor notions of bullshit and withholding information seem to be quiet accurate though not many tried to define these notions.

A problem which definitions based on modal logic have in common, is that they are coupled to this rather complicated form of logic with different operators which are usually not used in the application context of agent design. Therefore, more work is needed to transfer the results to actual agents, possibly by adapting simpler operators. This step might of course  damage the precision of the definitions but may help to apply them in practice. 
Furthermore, research is needed to come up with logically proofed methods to reason about when to use lies and how to recognise them. Sakama's considerations in \cite{Sakama2010a} about the preferred way of using dishonesty by deviating as little as possible from the truth are a first step in this direction. On the other hand, the approaches introduced in the next chapter might be more useful for practical applications after all, as they already embed formal definitions within agent communication and argumentation.

\section{Other approaches to formal definitions}

This chapter will examine other approaches and methods to define lies, deception and other notions. These are mainly embedded in existing agent architecture and communication systems bringing the advantage of being in the application context already. Nonetheless, these definitions can be written in a formal way to make them relevant to this review and to allow comparisons with the previously presented definitions with modal logic.

We will examine three examples to see how the definition can be embedded at very different levels of abstraction. The first example will look at a low-level approach operating on the belief base of an agent. The second paper uses a mid-level definition operating on speech acts and thirdly, a high-level definition will be given which is embedded in abstract argumentation frameworks.

Subsequently, we will begin to look at a low-level approach before continuing with higher level definitions.

\subsubsection{Low-level approaches}

An intelligent agent usually has a continuous update cycle of sensing the environment, updating the current beliefs about the world, deciding what to do by using desires and intentions and acting according to a plan which might achieve the current intention (following the idea first introduced by M.\,Bratman in \cite{Bratman1987}). The belief base of the agent which is updated in every cycle is the starting point of the low-level approaches. This way they allow for similar flexibility as modal logic while still being embedded in the agent's sense-decide-act cycle.

One example we won't examine in detail is given by F.\,De Rosis et al. in \cite{DeRosis2003}, where they employ a probabilistic model to model the belief base and define lies as a set of conditional probabilities thereupon.

We will look in more detail at a presentation of C.\,Sakama of 2012 \cite{Sakama2012b} which is based on a previous paper of Sakama et al. \cite{Sakama2011b} where they introduce 'logic programming' as a way to represent the knowledge base of an agent which allows to include disinformation. By disinformation they refer to all possible lies or bullshit given a certain agent's belief set.

The simplified version presented here uses the pair $\langle K,D\rangle$ as the belief base of an agent, such that K is the agent's knowledge and D contains all propositions that count as lies or bullshit according to equation  \ref{sa2012_lb}.
\begin{equation} \label{sa2012_lb}
\forall l\in D,K\vDash \neg l\vee(K\nvDash l\wedge K\nvDash\neg l)
\end{equation}
Additionally, the agent has the explicit goal to propose $g$ although it is not in the knowledge base ($K\nvDash g$) or to prevent $g$ from being proposed although it is in the knowledge base ($K\vDash g$). For this purpose, the agent can use an adapted knowledge base $(K\setminus J)\cup I$, where $J\subseteq K$ and $I\subseteq D$. Putting it more simply, the agent can ignore some parts of the knowledge base counteracting his intentions ($J$) and add some disinformation ($I$) enabling him to achieve his goal.

Lies, bullshit and withholding information are now defined using these sets $J$ and $I$ (Eq. \ref{sa2012_l} to \ref{sa2012_wi}).
\begin{align}
\label{sa2012_l}
\text{Lie, if }&I\neq\emptyset\wedge K\vDash \neg l \text{ for some }l\in I \\
\label{sa2012_b}
\text{Bullshit, if }&I\neq\emptyset\wedge K\nvDash \neg l \text{ for any }l\in I \\
\label{sa2012_wi}
\text{Withholding information, if }&I=\emptyset
\end{align}
Intuitively, lying is adding some additional information of which at least some is believed to be false. Bullshitting is adding some additional information of which none is believed to be false (nor to be true, as this would have enabled the agent to use the original set $K$ without adding anything). Finally, withholding information is leaving some information out without adding any disinformation.

The statement, addressee and intent-to-deceive conditions are not included explicitly in the definition. However, the framework implicitly includes all three conditions as the whole process is aimed at using the modified believe set to communicate $p$ or prevent $p$ from being communicated to someone with the intent to deceive (since communicating $p$ would not have been possible with the original belief $K$).

The definition of lying includes the believe-false condition and no notion of success and therefore fulfils all criteria. The same holds for the bullshit definition as it contains the indifferent believe and no notion of success either. Finally, withholding information fulfils all criteria as well, in this case by not communicating essential parts.

In his presentation, Sakama also gives some behavioural rules when to use each of the possibilities, using the preference ordering he already introduced in his modal logic papers of truth over withholding information over bullshit over lies.

Given that he manages to give definitions complying with all necessary criteria and nonetheless being incorporated in an actual agent design, this approach is one that should be taken into account.

In summary, the low-level approaches share the problem of either leaving out necessary conditions or including them only implicitly through the framework they are used in. However, if this implicit inclusion is acknowledged, at least the approach of Sakama in \cite{Sakama2012b} fulfils all criteria and, moreover, adds rules of how to apply the dishonesties in communication.

\subsubsection{Mid-level approaches}

The next level of abstraction is reached when the agents need to communicate with one another in a multi-agent system. The communication can be of various kinds, e.g. deliberation, inquiries, negotiation, persuasion or info-seeking as defined by Walter and Krabbe in \cite{Walton1995}. The constituting parts of communication are speech acts each consisting of a performative and the content, where the performative denotes the kind of speech (like requests, promises or assertions, see the classification of J.\,Searle \cite{Searle1976}). These speech acts, which are usually defined in a generally accepted language like FIPA-ACL \cite{FIPA2002}, can be modified to accommodate for lies and deception.

The paper of E.\,Sklar et al. \cite{Sklar2005}, we will look at in more detail, uses a speech-act based approach to integrate lies in agent communication. The main problem they face is the need to use the existing agent communication languages which are designed for truthful communication as pointed out by Parsons and Wooldridge in \cite{Parsons2003}. These existing languages allow a finite list of defined performatives for speech act, that is why no new performative 'lying' could be added. As Sklar et al. argue, this would not make sense anyway as the performatives are public and an agent who is publicly announcing that he is lying cannot really lie. 

What they are doing instead is to alter the existing pre- and postconditions of the performative 'assert' to allow an agent to make false assertions (i.e. to lie). The original definition of 'assert' is given in equation \ref{sk2005} (based on \cite{Parsons2003}).
\begin{equation} \label{sk2005}
\begin{split} 
\text{Locution: }&i\rightarrow j : assert(p) \\
\text{Pre-conditions: }&(S,p)\in\underline{S}(\Sigma_i\cup{CS}_j) \\
\text{Post-conditions: }&{CS}_{i,t+1}={CS}_{i,t}\cup\lbrace p\rbrace
\end{split}
\end{equation}
It symbolises that $i$ can make the assertion $p$ to agent $j$ when the argument for $p$ including its support $S$ can be drawn from the set of acceptable arguments $\underline{S}$ of all arguments in the knowledge base of I ($\Sigma_i$) combined with the already publicly uttered arguments of $j$ (${CS}_j$). Afterwards, the publicly uttered arguments of $i$ are updated with $p$ for the next iteration (Post-condition).
Based on this, Sklar et al. define a lie as the 'assert' speech act with the following pre- and postconditions in equation \ref{sk2005_l}.
\begin{equation} \label{sk2005_l}
\begin{split} 
\text{Locution: }&i\rightarrow j : assert(p) \\
\text{Pre-conditions: }&(S,\neg p)\in\underline{S}(\Sigma_i\cup{CS}_j)\text{ AND}\\
&(S',p)\in\underline{S}(\Sigma_i\cup{CS}_j\cup J_i) \\
\text{Post-conditions: }&{CS}_{i,t+1}={CS}_{i,t}\cup\lbrace p\rbrace
\end{split}
\end{equation}
The modified pre-conditions show that the contrary argument $\neg p$ is possible given only the knowledge base of $i$ and the publicly uttered commitments of $j$. However, by adding additional arguments $J_i$ which are not in the knowledge base of the agent $i$ originally, he can argue for $p$. 

This definition clearly contains the statement condition (by using the 'assert' speech act), it contains an addressee ($j$) and the believe-false condition ($(S,\neg p)\in\underline{S}(\Sigma_i\cup{CS}_j)$). However, the intention to deceive is missing.

The authors indicate that the additionally used knowledge $J_i$ needs to be remembered and maintained for each communication partner to ensure that the agent does not contradict itself. Furthermore, they think about possible applications, e.g. that lying might be an easier option when two different arguments are possible, one truthful but very complicated argument and one easy but dishonest argument. Moreover, they imagine the application area of negotiation where agents want to lie about their personal value of goods.

All in all, speech acts allow for a higher level of abstraction and at the same time for good definitions as well. The problem that occurred in the paper of Sklar et al. and which might be a general problem for this approach is that definitions are speaker-focused and do not account for intentions to deceive or even deception which is solely focused on the addressee.

\subsubsection{High-level approaches}

Still another level of abstraction can be achieved by constructing abstract argumentation frameworks as introduced by Dung in \cite{Dung1995}. They have the advantage of providing an easy problem understanding as they are able to depict the situation graphically. On the other hand, we will see that the high level of abstraction does not allow for unambiguous, clear definitions.

The abstract argumentation framework $\langle\mathcal{A},\rightharpoonup\rangle$ contains a set of arguments $\mathcal{A}$ and a defeat relation $\rightharpoonup$ between them. If one argument defeats another, it contradicts either the original argument itself or one of its supportive propositions. These argumentation frameworks are usually used to easily construct the set of arguments that should be accepted, in the sense that this acceptable set is consistent and does not contradict itself. Several measures can be applied for this purpose (e.g. a grounded semantic containing only the minimal set of arguments that need to be accepted in any case). When applying this approach to agent communication, each agent might propose its own set of acceptable arguments determined by their own knowledge. All publicly announced arguments form a new framework where the overall accepted arguments can be determined. Agents might have preferences over the finally accepted arguments, which in turn are useful when analysing the argumentation game-theoretically. Based on the argumentation frameworks, it is possible to define lies or especially withholding information as agents might want to improve their utility by influencing the acceptable arguments by adding additional arguments or hiding some arguments to break defeat chains. 

The paper we will examine here is by I.\,Rahwan et al. \cite{Rahwan2009}. An almost identical approach has also been described in their previous paper in \cite{Rahwan2008}. Formally, each agent $i$ in this paper has a true type $\mathcal{A}_i$ reflecting the true set of acceptable arguments given the agent's knowledge and a semantic. 
However, the publicly revealed type $\mathcal{A}_i^\star$ might be different from $\mathcal{A}_i$. (Thoroughly honest agents always have $\mathcal{A}_i^\star=\mathcal{A}_i$). Additionally, each agent has a utility function $u_i(\mathcal{A}_1^\star,\mathcal{A}_2^\star,\dotsc,\mathcal{A}_i^\star,\dotsc)$ for the possible final outcomes determined by the revealed types of all participants.

Based on this, the authors define their dishonest notions as in equations \ref{ra2008_1} and \ref{ra2008_2}.
\begin{gather}
\label{ra2008_1}
\text{Lying: }\mathcal{A}_i^\star\nsubseteq\mathcal{A}_i\\
\label{ra2008_2}
\begin{split}
\text{With. Inf.: }\mathcal{A}_i^\star&\subset\mathcal{A}_i \\ 
\text{AND }u_i(\mathcal{A}_1^\star,\mathcal{A}_2^\star,\dotsc,\mathcal{A}_i^\star,\dotsc)&>u_i(\mathcal{A}_1^\star,\mathcal{A}_2^\star,\dotsc,\mathcal{A}_i,\dotsc)
\end{split}
\end{gather}
It should be noted that the second condition for withholding information (the comparison between the utility functions) was not given formally, but in words.

Both definitions implicitly include the statement and the addressee condition, since the evaluation of the utility function is based on the preceding communicative act of each agent. One might also argue that the intention to deceive is included as well, because the other agents need to accept the wrongly revealed arguments which presumably is the intention of the agent. However, the believe-false condition for lying is not fulfilled as 
$\mathcal{A}_i^\star$ only needs to contain additional arguments about which nothing is known. Consequently, the lying definition might cover bullshit and lying possibly even paired with withholding information. The same problem holds for the withholding information definition in equation \ref{ra2008_2} as the higher utility does not suffice to show that the agent is inducing knowledge it does not believe. The definition would fit better to half-truths, as a true part of the knowledge is revealed with the intention to increase one's utility.

In the rest of the paper, the authors show that agents might have an incentive to lie as a Nash equilibrium for the argumentation game does not necessarily lead to honest behaviour. They conclude by proposing that this should be avoided by using restrictive rules. Another nice example for this game-theoretical analysis is shown in Rahwan et. al.'s subsequent paper \cite{Rahwan2009b}. Also, other authors like M.\,Caminada try to use abstract argumentation framworks to define lies and deception, but without giving formal definitions \cite{Caminada2009}. 

Altogether, this high level approach showed to be highly ambiguous when it comes to check the philosophical conditions. However, the practical implications for discussions and argumentation can be derived and investigated more easily due to the high abstraction.

\subsubsection{Summary}

Summarising the other approaches based on different level of agent communication, we can see some common problems. Embedding definitions in a context often either requires to assume implicitly some conditions or to leave out some conditions completely. Moreover, the definitions are fixed on the speaker, which is why no formal definition for deception was given. One the positive side, we can note the high application-orientation of all approaches which has been proved by applying the respective definitions to sensible examples. Moreover, higher levels of abstraction allow for high-level reasoning which might be impossible to do with a low-level logic-only approach. Given that agent technology will develop further, this area of embedding dishonesty in existing frameworks will surely become more important than the clear-cut modal logic definitions. 

\section{Open issues and conclusion}

In this review, we examined formal-logical definitions of lies, deception and other notions. Using the philosophical literature, we extracted guidelines that helped us to check whether formal definitions are correct and complete. Primarily, we looked at purely logical definitions and found some complying with all or almost all necessary conditions. We also assessed a few examples of other approaches which were formally defined but not based on pure logic. These were not as clear-cut and obviously correct, but had the great advantage of being embedded in an application environment. On the other hand, the modal logic definitions were accurate and concise (at least some), but not trouble-free applicable to applications.

In the remaining part we will look at the open issues that arise from this basis. At first, we start by listing some issues identified by the authors of the various papers, before adding other additional issues.

One question that seems to unanswered, as almost without exception all authors mentioned in chapter 4 referred to it, is how to include lying and deception successfully in agent communication and argumentation. Although some tried, as shown in chapter 4, there is yet no widely accepted solution that would allow to use the notion of lies or deception in practice.

Strongly connected is the question of how to detect lies or deception in communication (e.g. pointed out by Sakama in \cite{Sakama2011b}). There are papers considering this question, e.g. in \cite{Jones2006}, but not in a formal way which would allow to apply some detection algorithms.

A probably even more useful question is how to prevent agents from lying or trying to deceive \cite{Sakama2010b,Rahwan2009}. This connects to the question of how to design agent communication mechanisms that provide no incentive to lie, thereby enforcing honesty. Similarly, such questions have been considered in an informal way \cite{Ahern2012} but not applied formally.

Other open questions that have been identified by others are for example the influence of lies or bullshit arguments in discussions and if this could lead to some form of collective irrationality \cite{Caminada2009}. Or which computational complexity does reasoning about a strategy to lie or deceive actually have and if this is a hindrance to adopt such strategies in practice \cite{Rahwan2009}.

An issue which has not been addressed by any of the authors is the trade-off between simplicity/applicability and correctness. All authors using modal logic tried to use as complicated operators as necessary to define lies as correct as possible, whereas authors who included the definition in agent design used operators as simple as possible sacrificing correctness and conciseness. Possible solutions might need to adopt simpler operators which are usable in actual agents to define the notions as correctly as modal operators allowed. Besides simplification, unification of different ideas to one definition instead of many different ones might also help to apply the result in practice.

A second point which has not been addressed widely is how to react on detected lies or attempted deception. Although this step naturally succeeds the still open detection question, it is highly important as it alters the way agents communicate with another. They could introduce some kind of punishment or even ignore the liar in future encounters. The kind of expected punishment also influences the decision to lie as it changes the expected utility.

All in all, we can see that there are a number of open issues and possible improvements to be addressed. One can hope that these issues can be resolved to have, one day, independent self-acting agents confidently using lies, bullshit, deception or other notions avoiding detection, while at the same time being able to detect dishonesties of others in an environment that encourages honest behaviour and has clear rules on how to react to dishonest agents.

\bibliographystyle{acm}

\end{document}